\documentclass[preprint,showpacs,preprintnumbers,amsmath,amssymb]{revtex4}

\usepackage{graphicx}
\usepackage{dcolumn}
\usepackage{bm}

\begin{document}

\title{Fermionic Zero Modes in Gauge and Gravity Backgrounds on a
Sphere}

\author{Yu-Xiao Liu$^1$}\thanks{Corresponding author}\email{liuyx@lzu.edu.cn}
\author{Li-Jie Zhang$^2$}
\author{Yi-Shi Duan$^1$}

\affiliation{$^1$Institute of Theoretical Physics, Lanzhou
University, Lanzhou 730000, China}

\affiliation{$^2$Department of Mathematics and Physics, Dalian
Jiaotong University, Dalian 116028, China}

\begin{abstract}
In this letter we study fermionic zero modes in gauge and gravity
backgrounds taking a two dimensional compact manifold $S^2$ as
extra dimensions. The result is that there exist massless Dirac
fermions which have normalizable zero modes under quite general
assumptions about these backgrounds on the bulk. Several special
cases of gauge background on the sphere are discussed and some
simple fermionic zero modes are obtained.
\end{abstract}


\maketitle


Recently, there has been considerable activity in the study of
models that involve new extra dimensions
\cite{AntoniadisPLB,ADD,RS,Garriga}. The possible existence of
such dimensions got strong motivation from theories that try to
incorporate gravity and  gauge interactions in a unique scheme, in
a reliable manner. The idea dates back to the 1920's,  to the
works of Kaluza and Klein \cite{KK} who tried to unify
electromagnetism with Einstein gravity by assuming that the photon
originates from the fifth component of the metric.

An interesting topic is localization of different fields on a
brane. It has been shown that the graviton \cite{RS} and the
massless scalar field \cite{BajcB} have normalizable zero modes on
branes of different types, that the Abelian vector fields are not
localized in the RS model in five dimensions but can be localized
in some higher-dimensional generalizations of it \cite{OdaI}. In
Ref. \cite{RandjbarDaemiS} it was shown that there exist massless
Dirac fermions under quite general assumptions about the geometry
and topology of the internal manifold of the higher-dimensional
warp factor compactification. However, these fermionic modes are
generically non-normalizable. To avoid this problem, the authors
include a Yukawa-type coupling to a scalar field of a domain-wall
type. In Ref. \cite{Popov2005}, quiver gauge theory of nonabelian
vortices and noncommutative and instantons in higher dimensions
was studied and explicit zero modes of the Dirac operator on $S^2$
with a monopole bundle was obtained.

In our work \cite{WangYQ}, we study fermionic zero modes in the
background of self-dual vortex on a two-dimensional non-compact
extra space in 5+1 dimensions. In the Abelian Higgs model, we
present an unified description of the topological and
non-topological self-dual vortex on the extra two dimensions.
Based on it, we study localization of bulk fermions on a brane
with inclusion of Yang-Mills and gravity backgrounds in six
dimensions. Through two simple cases, it is shown that the vortex
background contributes a phase shift to the fermionic zero mode,
this phase is actually origin from the Aharonov-Bohm effect.

In this letter we consider fermionic zero modes in gauge and
gravity backgrounds taking a two dimensional sphere as extra
dimensions. The paper is organized as follows: Firstly, we
present Dirac equation and the effective Lagrangian in gauge and
gravity backgrounds. Secondly, we discuss and solve the fermionic
zero modes under several simple assumptions for gauge field on the
sphere. Thirdly, a brief conclusion is presented.


We shall consider a six-dimensional manifold $M^{4}\times S^{2}$
with the geometry
\begin{eqnarray}\label{Metric}
 ds^2 &=& G_{MN}dx^M dx^N \nonumber \\
      &=& g_{\mu \nu} dx^\mu dx^\nu
      - R^2 (d \theta^2 + \sin^2 \theta d \varphi^2),
\end{eqnarray}
where $g_{\mu \nu}$ is the four dimensional metric of the
space-time manifold $M^4$, capital Latin indices $M,N=0,\cdots,5$,
Greek indices $\mu,\nu=0,\cdots,3$, $x^4=\theta$ and $x^5=\varphi$
are the coordinates on the sphere $S^2$. The six-dimensional Dirac
equation is
\begin{equation}\label{DiracEq1}
\Gamma^{A}E^{M}_{A}(\partial_{M}-\Omega_{M}+ i
A_{M})\Psi(x,\theta,\varphi)=0,
\end{equation}
where $E^{M}_{A}$ is the {\sl sechsbein} with
\begin{equation}\label{sechsbein}
E^{M}_{A} = \left ( e^{\mu}_{a}
\delta^{a}_{A},\frac{1}{R}\delta^{4}_{A},\frac{1}{R
\sin\theta}\delta^{5}_{A} \right ),
\end{equation}
and capital Latin indices $A,B=0,\cdots,5$ correspond to the flat
tangent six-dimensional Minkowski space, $\Omega_M=\frac{1}{4}
\Omega_M^{AB}\Gamma_{A}\Gamma_{B}$ is the spin connection, where
Dirac matrices $\Gamma^A$ follow the relation
$\Gamma^A\Gamma^B+\Gamma^B\Gamma^A=2\eta^{AB}I$.

The non-vanishing components of $\Omega_M$ are
\begin{equation}
\Omega_{\mu}=\omega_{\mu},\;\;\; \Omega_{5}=\frac{1}{2} \cos\theta
\Gamma^4 \Gamma^5,
\end{equation}
where $\omega_{\mu}=\frac{1}{4}
\omega_{\mu}^{ab}\gamma_{a}\gamma_{b}$ is the spin connection
derived from the vielbein $e_{\mu}^{a}$ with
$g_{\mu\nu}=e_{\mu}^{a} e_{\nu}^{b} \eta_{ab}$, lower case Latin
indices $a,b=0,\cdots,3$ correspond to the flat tangent
four-dimensional Minkowski space.

Assume $A_{\mu}=A_{\mu}(x)$,
$A_{\theta}=A_{\theta}(\theta,\varphi)$,
$A_{\varphi}=A_{\varphi}(\theta,\varphi)$. The six-dimensional
Dirac equation (\ref{DiracEq1}) then becomes
\begin{eqnarray}
 &&\left \{ D_M + \frac{1}{R} \Gamma^4 \left(\partial_{\theta}
        - \frac{1}{2} \cot\theta + i A_{\theta} \right) \right. \nonumber \\
  &&~~+ \left. \frac{1}{R \sin\theta} \Gamma^5(\partial_{\varphi}
        + i A_{\varphi}) \right \} \Psi = 0.
        \label{DiracEq2}
\end{eqnarray}
where
\begin{eqnarray}
 D_M=\Gamma^a e^{\mu}_{a} (\partial_{\mu} - \omega_{\mu}
        + i A_{\mu})
\end{eqnarray}
We denote the Dirac operator on a sphere $S^2$ with $D_S$:
\begin{equation} \label{DiracOperator}
D_S =\frac{\bar{\Gamma}}{R} \left \{ \Gamma^4(\partial_{\theta} -
\frac{1}{2} \cot\theta + i A_{\theta}) + \frac{1}{\sin\theta}
\Gamma^5(\partial_{\varphi} + i A_{\varphi}) \right \},
\end{equation}
where $\bar{\Gamma}=\Gamma^0 \Gamma ^1 \Gamma ^2 \Gamma ^3,$ and
expand any spinor $\Psi$ in a set of eigenvectors $\phi_m$ of this
operator \begin{equation}   D_S\phi_m=\lambda_m\phi_m.
  \end{equation} There may exist a set of discrete eigenvalues $\lambda_m$ with
some separation. All these eigenvalues play a role of the mass of
the corresponding four-dimensional excitations \cite{Libanov}. We
assume that the energy scales probed by a four-dimensional
observer are smaller than the separation, and thus even the first
non-zero level is not excited. So, we are interested only in the
zero modes of $D_S$: \begin{equation}   D_S\phi=0.
\end{equation} This is just the Dirac equation on the manifold
$S^2$ with gauge and gravity backgrounds. For fermionic zero
modes, we can write \begin{equation}
\Psi(x,\theta,\varphi)=\psi(x) \phi(\theta,\varphi),
\end{equation}
where $\phi$ satisfy $D_S\phi(\theta,\varphi)=0$. The effective
Lagrangian for $\psi$ then becomes
\begin{eqnarray}
L_{eff}&=&\int d\theta d\varphi \sqrt{-G}
\bar{\Psi} \Gamma^A E^M_A (\partial_M - \Omega_M + i A_M) \Psi \nonumber\\
&=& L_{\psi} \times  R^2\int d\theta d\varphi \sin\theta \phi^\dag
\phi,
\end{eqnarray}
where
\begin{eqnarray}
L_{\psi}= \sqrt{-g} \bar{\psi} \Gamma^a e^{\mu}_a (\partial_{\mu}
- \omega_{\mu} + i A_{\mu}) \psi.
\end{eqnarray}
Thus, to have the localization of gravity and finite kinetic
energy for $\psi$, the above integral must be finite. This can be
achieved if the function $\phi(\theta,\varphi)$ does not diverge
on the whole sphere $S^2$. So, for any regular solutions, the bulk
fermions can be localized on a brane with the use of gravity and
gauge fields. This is very different from the case considered by
Randjbar-Daemi and Shaposhnikov \cite{RandjbarDaemiS}, who did
this work on a $D=D_1 + D_2 +1$ dimensional manifold with the
geometry
\begin{equation} ds^2=e^{A(r)}\eta_{\mu\nu}dx^{\mu}dx^{\nu} +
e^{B(r)}g_{mn}dy^m dy^n + dr^2,  \end{equation} under the
assumptions $A_{\mu}=A_r=0$ and $A_m=A_m(y)$. Their effective
Lagrangian for $\psi$ is
\begin{eqnarray}
&&\int
dr dy \sqrt{-G} \bar{\Psi} \Gamma^A E^M_A (\partial_M - \Omega_M + i A_M) \Psi \nonumber\\
&=& \bar{\psi} \Gamma^a \delta^{\mu}_a \partial_{\mu} \psi \int dr
e^{-A(r)/2} \int dy \sqrt{det(g_{mn})} \phi^\dag \phi,~~~
\end{eqnarray}
and the difficulty is that the integral $\int dr e^{-A(r)/2}$ is
infinite for the exponential warp-factor $A \propto -|r|$
considered in the literature so far.


In the following, we discuss and solve the fermionic zero modes
under some simple assumptions for $g_{\mu\nu}$ and gauge field
$A_{\theta}$ and $A_{\varphi}$ on the sphere.


Case I:  $A_{\varphi}=A(\varphi)$, $A_{\theta}=A(\theta)$.

The Dirac equation (\ref{DiracEq2}) then becomes
\begin{eqnarray}
 &&\left\{ D_M + \frac{1}{R} \Gamma^4 \left( \partial_{\theta}
       - \frac{1}{2} \cot\theta + i A(\theta) \right) \right. \nonumber \\
 && ~~+ \left. \frac{1}{R \sin\theta} \Gamma^5(\partial_{\varphi}
       + i A(\varphi)) \right \} \Psi = 0.
\end{eqnarray}
Here $\Psi$ can be written as the following form:
\begin{equation}
\Psi (x,\theta,\varphi)=\psi (x) f(\theta) h(\varphi),
\end{equation}
where $f$ and $h$ satisfy
\begin{eqnarray}
\left ( \partial_{\theta} - \frac{1}{2} \cot\theta + i A(\theta)
\right )
f(\theta) & = 0,\\
(\partial_{\varphi} + i A(\varphi)) h(\varphi) & = 0.
\end{eqnarray}
Solve the last two equations one can easily get the formalized
solutions:
\begin{eqnarray}
 f(\theta) &=& \sqrt{\sin\theta}~ e^{-i\int{d \theta A(\theta)}},\\
 h(\varphi) &=& e^{-i \int{d \varphi A(\varphi) }},
\end{eqnarray}
The effective Lagrangian for $\psi$ then becomes
\begin{eqnarray}
L_{eff} = L_{\psi} &\times& R^2 \int^{\pi}_{0} {d\theta
\sin^2\theta \| e^{-i\int{d\theta A(\theta)}} \|^2} \nonumber \\
&\times&\int^{2\pi}_{0}{d\varphi}\| e^{-i\int{d\varphi A(\varphi)
}} \|^2.
\end{eqnarray}
If $A(\theta)$ and $A(\varphi)$ are real functions, then the
effective Lagrangian can be simplified as
\begin{eqnarray}
L_{eff}= \pi^{2}R^2 L_{\psi}.
\end{eqnarray}
This result shows that, whatever the forms of $A_{\theta}(\theta)$
and $A_{\varphi}(\varphi)$ are, the effective Lagrangian for
$\psi$ has the same form. But if the gauge fields are not real
functions, they would have a effect on the effective Lagrangian,
and the integral may not be convergent.


Case II: $A_{\varphi}=A(\theta)$, $A_{\theta}=0$.

This simple case is considered by J. M. Frere etc. \cite{Frere}.
But a complex scalar field $\Phi(\theta,\varphi)$ on $S^2$ is
included in their work. For our current ansatz, the Dirac equation
is
\begin{eqnarray}
 &&\left \{ D_M
       + \frac{1}{R} \Gamma^4 \left( \partial_{\theta}
       - \frac{1}{\sin\theta}\Gamma^4 \Gamma^5 A(\theta)\right. \right. \nonumber \\
&&~~-\left.\left. \frac{1}{2} \cot\theta
       \right)
       +  \frac{1}{R \sin\theta} \Gamma^5\partial_{\varphi}
       \right \} \Psi = 0,
\end{eqnarray}
with
\begin{equation}
\Psi(x,\theta,\varphi)=\psi(x) f(\theta),
\end{equation}
where $f$ satisfies
\begin{equation}
\left ( \partial_{\theta} - \frac{1}{2} \cot\theta
-\frac{1}{\sin\theta}\Gamma^4 \Gamma^5 A(\theta) \right )
f(\theta)=0.\label{f}
\end{equation}
Let $f(\theta)=\left(f_{1}(\theta),f_{2}(\theta)\right)^{T}$, and
imposing the chirality condition $\gamma^5
f_{i}(\theta)=+f_{i}(\theta)$, the above equation can be written
as
\begin{equation}
\left ( \partial_{\theta} - \frac{1}{2} \cot\theta -i
\frac{A(\theta)}{\sin\theta} \right ) f_{i}(\theta)=0.
\end{equation}
So, the formalized solution of $f$ can be given out
\begin{equation}
f_{1,2}(\theta) =\sqrt{\sin\theta}~ e^{-\int{d\theta
\frac{A(\theta)}{\sin\theta} }}
\left(%
\begin{array}{c}
  1 \\
  1 \\
\end{array}%
\right) \otimes
\left(%
\begin{array}{c}
  1 \\
  0 \\
\end{array}%
\right).
\end{equation}
The effective Lagrangian for $\psi$ then becomes
\begin{eqnarray}
L_{eff}\propto L_{\psi} \times  \int^{\pi}_{0} {d\theta
\sin^2\theta \| e^{-\int{d\theta \frac{A(\theta)}{\sin\theta}
}}\|^2}.
\end{eqnarray}
This result shows that, if $A(\theta)$ is a pure imaginary
function, then the effective Lagrangian for $\psi$ has the same
form and the extra dimension contributes a constant to the
Lagrangian.


Now we come to the issue of the presence of the zero modes. The
the number of zero-modes of the Dirac operator is decided by the
index of it. The index of the Dirac operator on manifold $K$ is
defined as the difference $n_{+} - n_{-}$ between the number
$n_{+}$ of right-handed four-dimensional fermions obtained by
dimensional reduction and the number $n_{-}$ of left-handed 4D
fermions. This number is a topological quantity of the manifold
upon compactification and the gauge bundles the Dirac operator
might be coupled to. Indeed, this index can be computed in terms
of characteristic classes of the tangent and gauge bundles. The
Atiyah--Singer index theorem in two dimensions gives the
difference \cite{Atiyah1968_485, Atiyah1968_546, Atiyah1973_279}
\begin{equation}
n_{+} - n_{-}=\frac{e}{4\pi} \int_\mathcal{M} d^2 q \,
\varepsilon^{\mu\nu} F_{\mu\nu} \,,
\end{equation}
where $\varepsilon^{\mu\nu}$ $(\varepsilon^{12}=1)$ is the
contravariant Levi--Civita tensor {\em density} in two dimensions,
and $F_{\mu\nu}$ is the field strength of $A_{\mu}$,
\begin{equation}
F_{\mu\nu}\equiv \partial_{\mu}A_{\nu}-\partial_{\nu}A_{\mu}
-ie[A_{\mu}, A_{\nu}] \,.
\end{equation}
If we take $K=S^2$ with a U(1) magnetic monopole field of charge
$n$ on it, the number of chiral families will then be equal to $n$
\cite{Randjbar1983,Deguchi2005}. Here, we can consider zero modes
of the Dirac operator in the background of Abelian gauge
potentials representing Dirac strings and center vortices on the
torus $T^2$. The result is for a two-vortex gauge potential
(smeared out vortices) there is one normalizable zero mode which
has exactly one zero on the torus \cite{Reinhardt2002}. The
probability density of the spinor field is peaked at the positions
of the vortices.

In conclusion, we studied fermionic zero modes and effective
Lagrangian in gauge and gravity backgrounds taking a two
dimensional compact manifold $S^2$ as extra dimensions. The result
is that there exist massless Dirac fermions under quite general
assumptions about the gauge field on the bulk. These fermionic
zero modes are generically normalizable and we need not include
any other field. Furthermore, Several special cases of these
backgrounds are also discussed. Especially, in the case of
$A_{\theta} = 0$ and $A_{\varphi} =A_{\varphi}(\theta)$, we got
very simple fermionic zero mode on the sphere and the effective
Lagrangian for $\psi$, which has the same form whatever
$A_{\varphi}$ is. For the case of a torus and a string-like
defect, the corresponding solutions in gauge and self-dual vortex
backgrounds have studied in Ref. \cite{LiuYXMPLA2006} and Ref.
\cite{LiuYXNPB2007}, respectively.


This work was supported by the National Natural Science Foundation
of the People's Republic of China (No. 502-041016, No. 10475034
and No. 10705013) and the Fundamental Research Fund for Physics
and Mathematic of Lanzhou University (No. Lzu07002).


\begin{thebibliography}{99}

\bibitem{AntoniadisPLB}
 I. Antoniadis,
    Phys. Lett. \textbf{B 246} (1990) 377.
 I. Antoniadis, N. Arkani-Hamed, S. Dimopoulos and G. Dvali,
    Phys. Lett.  \textbf{B436} (1998) 257. 

\bibitem{ADD}
 N. Arkani-Hamed, S. Dimopoulos and G. Dvali,
    Phys. Lett. {\bf B429} (1998) 263.


\bibitem{RS}
 L. Randall and R. Sundrum,
 Phys. Rev. Lett. {\bf 83} (1999) 3370;
 Phys. Rev. Lett. {\bf 83} (1999) 4690.

\bibitem{Garriga}
 J. Garriga and T. Tanaka,
     Phys. Rev. Lett. {\bf 84} (2000) 2778;
 T. Shiromizu, K. I. Maeda and M. Sasaki,
     Phys. Rev.  {\bf D62} (2000) 024012;
 A. N. Aliev and A. E. Gumrukcuoglu,
    Class. Quant. Grav. {\bf 21} (2004) 5081, [arXiv:hep-th/0407095].


\bibitem{KK}
 T. Kaluza, Math. Phys. {\bf K1} (1921) 966;
 O. Klein, Z. Phys. {\bf 37} (1926) 895.

\bibitem{BajcB}
 B. Bajc and G. Gabadadze,
 Phys. Lett. {\bf B474} (2000) 282.

\bibitem{OdaI}
 I. Oda, Phys. Lett. {\bf B496} (2000) 113.

\bibitem{Popov2005}
 A. D. Popov and R. J. Szabo,
     J. Math. Phys. \textbf{47} (2006) 012306,
     [arXiv:hep-th/0504025].

\bibitem{RandjbarDaemiS}
 S. Randjbar-Daemi and M. Shaposhnikov,
    Phys. Lett. {\bf B492} (2000) 361;
 Y. X. Liu, L. Zhao and Y. S. Duan,
    JHEP \textbf{0704} (2007) 097,
    [arXiv:hep-th/0701010].


\bibitem{WangYQ}
 Y. Q. Wang, T. Y. Si, Y. X. Liu and Y. S. Duan,
 Mod. Phys. Lett. {\bf A20} (2005) 3045.

\bibitem{Libanov}
 M. V. Libanov and S. V. Troitsky,
 Nucl. Phys. {\bf B599} (2001) 319.

\bibitem{Frere}
  J. M. Frere, M. V. Libanov, E. Y. Nugaev and S, V. Troitsky,
  JHEP {\bf 0306} (2003) 009.

\bibitem{Atiyah1968_485}
 M. F. Atiyah and I. M. Singer,
 Ann. Math. {\bf 87} (1968) 485.

\bibitem{Atiyah1968_546}
 M. F. Atiyah and I, M. Singer,
 Ann. Math. {\bf 87} (1968) 546.

\bibitem{Atiyah1973_279}
 M. F. Atiyah, R. Bott and V. K. Patodi,
 Invent. Math. {\bf 19} (1973) 279.

\bibitem{Randjbar1983}
 S. Randjbar-Daemi, A. Salam and J. Strathdee,
 Nucl. Phys. {\bf B214} (1983) 491.

\bibitem{Deguchi2005}
 S. Deguchi and K. Kitsukawa,
 {\em Charge quantization conditions based on
      the Atiyah--Singer index theorem},
 arXiv:hep-th/0512063.

\bibitem{Reinhardt2002}
 H. Reinhardt, O. Schr\"{o}der, T. Tok and V. C. Zhukovsky,
 Phys. Rev. {\bf D66} (2002) 085004.

\bibitem{LiuYXMPLA2006}
 Y. S. Duan , Y. X. Liu and Y. Q. Wang,
 Mod. Phys. Lett. {\bf A21} (2006) 2019;
 Y. X. Liu, L. Zhao, L. J. Zhang and  Y. S. Duan,
     Int. J. Mod. Phys. \textbf{A22} (2007) 3643;
 Y. X. Liu, Y. Q. Wang and Y. S. Duan,
    {\em Fermionic zero modes in self-dual vortex background on a torus},
    Commun. Theor. Phys. to be published (2007).

\bibitem{LiuYXNPB2007}
 Y. X. Liu, L. Zhao, X. H. Zhang and Y. S. Duan,
    Nucl. Phys. \textbf{B785} (2007) 234, arXiv:0704.2812[hep-th].

\end{thebibliography}
\end{document}